\documentstyle[prl,multicol,aps,epsf]{revtex}

\begin{document}
\draft

\title{
Landau quantization and equatorial states on a surface of a nanosphere.
}
\author{ D.N. Aristov}
\address{
Petersburg Nuclear Physics Institute, Gatchina 188350 St.Petersburg,
Russia }
\date{\today}

\maketitle

\begin{abstract}
The Landau quantization for the electron gas on a surface of sphere is
considered. We show that in the regime of strong fields the lowest
energy states are those with magnetic quantum numbers $m$ of order
of $\Phi/\Phi_0$, the number of magnetic flux quanta piercing the
sphere. For the electron gas of low density (semiconducting
situation), it leads to the formation of the electronic stripe on the
equator of the sphere in high fields.
\end{abstract}


\begin{multicols}{2} \narrowtext

The electronic properties of cylindrical and spherical nanosize objects
attract much of the theoretical interest last years.  This interest is
mostly related to the physics of carbon macromolecules and,
particularly, to the transport properties of carbon nanotubes.
\cite{currents} One meets the spherical nanosize objects in the studies
of the nonlinear optical response in composite materials
\cite{composites}, of simple metal clusters \cite{deHeer} and of the
photonic crystals on the base of synthetic opals \cite{opals} . In the
most of these studies, the coating of the nano-sphere is characterized
by an effective dielectric function.  \cite{coated} It was noted
however \cite{Ekardt} that this approach should be revised if the
coating of a sphere has a width of a few monolayers, which limit is
allowed by modern technologies.

In the recent papers we considered the electron gas on a sphere.
We showed that various correlation functions in such gas exhibit maxima
when the electrons are at the points-antipodes (north and south poles).
\cite{sphereC} The exact solution was found for a problem in the
uniform magnetic field and the limits of weak and high fields were
investigated.  \cite{sphereM}
In the high-field regime the formation of Landau levels was shown.
The complexity of the special functions describing the exact solution,
however, complicates the analysis of the physical picture in the
high-field regime. In this paper we explore qualitative arguments,
supported by the numerical calculations, to clarify the issue.
We find that the minimum energy to the Hamiltonian is provided by the
electronic states located near the equator of the sphere.
For low densities of the gas one thus expects, that the high field
pushes the electrons towards the equator and forms an electronic ring
there.

We consider the electron gas moving within a thin layer on a
surface of the sphere of radius $r_0$.
We assume the Hamiltonian of the form
$     {\cal  H} = -\frac{\nabla^2}{2m_e} + U(r), $
with $m_e$ the electron mass.
The chemical potential $\mu$ defines the total number of electrons $N$
(with one projection of spin)
and the areal density $\nu = N/(4\pi r_0^2)$.
The confining quantum-well potential $U(r)$ restricts the radial motion
within the thin layer $\delta r \ll r_0$.
We are interested in the case
$\delta r < \nu ^{-1/2}$, when the first excited state of the radial
motion lies above the chemical potential. Then
one can ignore the radial component of the wave function and put
$r=r_0$ in the remaining angular part of the Hamiltonian.
In the absence of the magnetic field we have
        $
        {\cal H}_{\Omega}^{(0)}=- (2m_er_0^2)^{-1}\Delta_{\Omega}
        $
with $\Delta_\Omega$ being the angular part of the Laplacian.
The solutions to this Hamiltonian are the
spherical harmonics $Y_{lm}$ and the spectrum is that of a free rotator
model :

        \[ 
        \Psi^{(0)}(\theta,\phi)
        = r_0^{-1} Y_{lm}(\theta,\phi), \quad
        E_{l}^{(0)} = (2m_er_0^2)^{-1} l(l+1).
        \] 

 In the presence of the uniform magnetic field ${\bf B}$
directed towards a north pole of the sphere ($\theta =0$) we choose
the gauge of the vector potential as ${\bf A}=\frac12 ({\bf
B}\times{\bf r})$. Then the angular part ${\cal H}_\Omega$ of the
Hamiltonian ${\cal H} = \frac1{2m_e} (-i\nabla + e {\bf A})^2  + U(r)$
acquires the form

       \begin{equation}
       {\cal H}_\Omega=(2m_er_0^2)^{-1}
       [-\Delta_\Omega+
       2ip \frac{\partial}{\partial \phi}
       +p^2\sin^2\theta ]
       \label{spheq}
       \end{equation}
The ruling parameter here is
$p= {\pi Br_0^2}/{\Phi_0}$
with the magnetic flux quantum $\Phi_0 =2\cdot 10^ {-15}$
T$\cdot$m$ ^2$.
For a sphere of radius
$r_0=100\,$nm one has $p=1$ at the field $B\simeq 600\,$Oe.
The solutions of (\ref{spheq}) are given by the oblate
(angular) spheroidal functions and were analyzed in \cite{sphereM} to
some detail.

In the weak-field regime, $p\sim 1$, the jumps in the static magnetic
susceptibility $\chi$ at half-integer $p$ were demonstrated. The
amplitude of these jumps is parametrically larger than the Pauli spin
contribution and decreases with the increase of $p$. It was shown
that the weak-field regime ends at $p^2 \sim p_c^2 = 2\sqrt{N}$.
For $r_0\sim 100\,$nm (the case of opals) and in the
metallic situation, e.g., at the densities $\nu\sim
10^{14}\,$cm$^{-2}$, we have $N\sim 10^5$ and $p_c \simeq 30$.

On the other hand, at the lower (semiconducting) densities, $\nu \sim
10^{10}\,$cm$^{-2}$, we have $N\sim 10$. Formally in this case $p_c
\simeq 3$, i.e. the field is not small already at $p$ of order of
unity.  The jumps in the susceptibility were predicted in
\cite{sphereM} at the assumption $\sqrt{N}\gg1$ which is violated in
the latter case of lower $\nu$.  At the same time the experimentally
accessible fields of order of $6\,$T result in $p\sim 100$ for the
spheres with $r_0\sim 100\,$nm, i.e. we come into the strong field
regime.

For strong fields, $p\to \infty$, one observes the eventual formation
of the Landau levels (LL).  The spherical geometry brings into the
problem some peculiarities which were partly discussed in
\cite{sphereM}. First is the incomplete restructuring of the spectrum
into the LL scheme.  This restructuring takes place only for levels
with initial momentum $l$ lower than $p$, and the field remains weak
for the levels with $l>p^2$.  Secondly, the field-induced two-well
potential $p^2\sin^2\theta$ in (\ref{spheq}) localizes the electron
states with moderate magnetic quantum numbers $m$ ($|m|\ll p$) near the
poles $\theta=0$ and $\theta=\pi$. As a result, the correlations within
one hemisphere only survive.  Specifically, if an electron was
initially in the northern hemisphere, then the probability to find it
in the southern hemisphere is exponentially small.

At the same time, the spherical geometry produces yet another effect
which is to be discussed here. The effective potential in the
strong-field regime can be written as

        \begin{eqnarray}
        U_{eff}(\theta)
        &=& \frac{p\omega_c}{4}
        \left( \frac{m/p}{\sin\theta} +\sin\theta
        \right)^2
        \end{eqnarray}
with the cyclotron frequency $\omega_c = eB/m_e$.
At small negative $m$ we have two minima of
$U_{eff}(\theta)$ near $\theta_0 = \arcsin (\sqrt{|m|/p})$ and
$\theta_0 = \pi- \arcsin (\sqrt{|m|/p})$ where we expand

        \begin{equation}
        U_{eff}(\theta_0+\theta)
        \simeq (\omega_c p
        \cos^2\theta_0 )\,
        \theta^2
        \label{harmonic}
        \end{equation}
Rescaling here $\theta \to x /\sqrt{2p|\cos\theta_0|}$ we arrive at the
quantum oscillator problem of the form

\[{\cal H} \simeq
\frac{\omega_c|\cos\theta_0|}2 \left( -d^2/dx^2 + x^2
 \right),\]

\noindent
i.e. the well-known Landau quantization. As long as
$|\cos\theta_0|\sim1$, the wave-functions are extended at the scale
$|\theta - \theta_0| \sim p^{-1/2}$.  The possibility of quantum
tunneling between $\theta_0$ and $\pi-\theta_0$ produces the
exponentially small splitting between the states centered at these
points. \cite{sphereM}

Thus we see that the energy levels are labeled by two quantum numbers,
magnetic number $m$ and LL number $n$, with approximate
double degeneracy for given $m,n$.

This simple picture becomes inadequate, when
$|m|\simeq p$ and $\theta_0 \simeq
\pi/2$. In this case the harmonic potential in (\ref{harmonic})
weakens, which makes necessary the consideration of the fourth-order
terms in the expansion. We have in this case

        \begin{equation}
        U_{eff}(\theta+\pi/2)
        \simeq \frac{\omega_c p}4
        \theta^4,\quad  |m|=p
        \label{anharmonic}
        \end{equation}
Rescaling now $\theta \to x p^{-1/3}$ we arrive at the following
Schr\"odinger equation :

\[
\frac{\omega_c}{4p^{1/3}} \left( -d^2/dx^2 + x^4 \right)
\psi   =  E\psi ,\]

The solution of the last equation apparently is not known \cite{Kamke}.
For our purposes it suffices to note that the energy scales as
$\omega_c/p^{1/3} \ll \omega_c$ and the wave-functions on the
equator extend on the scale $|\theta - \pi/2| \sim p^{-1/3}$.
In addition, we have no situation with two-well potential now and
the energy levels are separated by the same scale
$\omega_c/p^{1/3}$. The crossover between Eq.(\ref{harmonic}) and
Eq.(\ref{anharmonic}) takes place at $1-|m|/p \sim p^{-2/3}$.

With the further increase of $|m|$,
at $|m|>p$, the minimum value of $U_{eff}$ is found at
$\theta=\pi/2$ and increases rapidly with $|m|$. In this case the
energy levels $E_{lm} \sim \omega_c (m+p)^2/p$ and thus
lie well above those with $|m| < p$.

\begin{figure}[tbp]
\centerline{\epsfxsize=8cm
\epsfbox{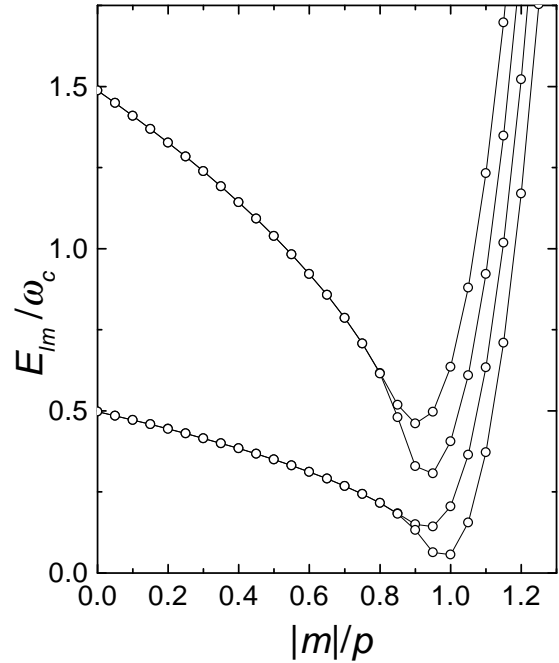}}
\caption{ The dependence of the four lowest energy levels
$E_{lm}$ on the magnetic number $m<0$  in the strong-field
regime, $p= \Phi/\Phi_0=100$.  The calculated points are shown by
circles, the lines are guide to the eye.
}
\end{figure}

We illustrate these qualitative results by the numerical calculations.
We found the spectrum of Eq.(\ref{spheq}) by diagonalizing ${\cal
H}_\Omega$ in the basis of Legendre polynomials
$P_{m+n}^m(\cos\theta)$ with $0\leq n \leq 200$ \cite{Flammer}.
The results are shown in the Fig.\ 1.
One verifies that the ``equatorial'' states with $|m|\simeq p
\gg1$ provide the minimum eigenvalues to the Hamiltonian.
This dependence of the energy level scheme on $m$ is probably of minor
importance, if we consider the situation of a metal, \cite{sphereM}
when the chemical potential lies well above the bottom of the
conduction band ($\mu \gg \omega_c$).
In this case the number of electrons $N$ on the sphere is expected to
exceed the number of flux quanta $p$. Then several Landau levels are
occupied (several lines in the Fig.\ 1) and the electrons are
distributed upon the whole sphere.

At the same time, the semiconducting coating of the sphere can lead to
a different result.
Indeed, if the cyclotron frequency $\omega_c$ is enough
high, then the ``polar'' states with small negative $m$ are poorly
occupied. Meanwhile, the ``equatorial'' states  with the energies $\sim
\omega_c p^{-1/3}$ are occupied to a larger extent.  This produces
the effective ring on the equator of the sphere.  The criterion for
this phenomenon is $N \alt p$ or, equivalently, $B \agt \Phi_0\nu$.
Note that the latter inequalities correspond to the partially filled
lowest Landau level in the usual planar geometry.

We see that in the spherical geometry of the electron gas the states
with higher $|m|$ possess the lower energy. Our situation is thus
opposite to the one discussed for the quantum Hall edge states.
\cite{qHes} Nevertheless both problems have a common ingredient, the
linear-in-$m$ spectrum for a given $n$ (in our case in two domains,
$|m|<p$ and $|m|>p$ ). Having effectively a case of one spatial
dimension, we can consider the interaction effects as well.  The
problem however has a certain subtlety which is described below.

In certain cases one may hope to ignore the interaction between
the states belonging to different ``Landau levels'' (different curves in
the Fig.\ 1). Considering now the lowest LL, one
observes familiar branches of right- and left-going fermions, $|m|\alt
p$ and $|m|\agt p$, with the negative and positive ``Fermi velocities''
$v_F = dE_{lm}/dm$, respectively. The absolute values of $v_F$ for
left and right movers are different. This point alone makes it
difficult to pass to a bosonization description with one scalar field
for both movers, and the notion of the chiral Luttinger liquid appears.
A thorough consideration of the latter problem is beyond the scope of
this study.

In conclusion, we considered the Landau quantization for the
electron gas on a surface of sphere. The exact solution of this
problem involves complicated functions, which are not very instructive
for the analysis of states with large magnetic numbers $m$ for the
electron motion. We elucidate the role of the ``equatorial'' states
with large $m$ both analitically and numerically. These states are
lower in energy, thus the electronic stripe on the equator can be
realized for the semiconducting coating of the sphere in high magnetic
fields.


I thank S.G. Romanov, A.G. Yashenkin, K. Hansen, V.A. Kulbachinskii for
useful discussions and communications.
The financial support from the Russian State Program for
Statistical Physics (Grant VIII-2), grant FTNS 99-1134 and grant
INTAS 97-1342 is gratefully acknowledged.


\end{multicols}

\end{document}